\documentclass[aps,pra,showpacs,twocolumn,groupedaddress]{revtex4}
\usepackage{graphicx,amsmath,amssymb}
\begin{document}
\newtheorem{Theorem}{Theorem}

\title{Quantum search algorithm by adiabatic evolution under a priori probability}

\author{Zhaohui Wei and Mingsheng Ying}

\affiliation{ State Key Laboratory of Intelligent Technology and
Systems, Department of Computer Science and Technology, Tsinghua
University, Beijing, China, 100084}

\begin{abstract}

Grover's algorithm is one of the most important quantum
algorithms, which performs the task of searching an unsorted
database without a priori probability. Recently the adiabatic
evolution has been used to design and reproduce quantum
algorithms, including Grover's algorithm. In this paper, we show
that quantum search algorithm by adiabatic evolution has two
properties that conventional quantum search algorithm doesn't
have. Firstly, we show that in the initial state of the algorithm
only the amplitude of the basis state corresponding to the
solution affects the running time of the algorithm, while other
amplitudes do not. Using this property, if we know a priori
probability about the location of the solution before search, we
can modify the adiabatic evolution to make the algorithm faster.
Secondly, we show that by a factor for the initial and finial
Hamiltonians we can reduce the running time of the algorithm
arbitrarily. Especially, we can reduce the running time of
adiabatic search algorithm to a constant time independent of the
size of the database. The second property can be extended to other
adiabatic algorithms.
\end{abstract}

\maketitle

Quantum search algorithm developed by Grover \cite{GROVER97} is
one of the main applications of quantum computation. If there is
an unsorted database with N items in which there is only one
marked item satisfies a given condition, then using Grover's
algorithm we will find the object in $O(\sqrt{N})$ steps instead
of $O(N)$ classical steps. When the quantum search algorithm was
proposed, it was implemented by a discrete sequence of unitary
logic gates, which is regarded as the standard paradigm for
quantum computing. Now a new quantum computing paradigm based on
quantum adiabatic evolution has been proposed \cite{FGGS00}. It
has been shown that the quantum search algorithm can be
implemented by the adiabatic evolution, which also gives rise to a
quadratic speed up as Grover's algorithm \cite{RC02}.

In Grover's algorithm and the local adiabatic search algorithm in
\cite{RC02}, before search we just know the number of marked items
in the unsorted database, and we don't know any further
information about the marked item. However, sometimes maybe we
know something about some possible locations of the marked item,
then we think that perhaps we can use this information to help
search. In this paper, we will show that if we know a priori
probability about the marked item, we can modify the quantum
adiabatic search algorithm to search faster. Then we will show
another property of adiabatic search algorithm that conventional
quantum search algorithm doesn't have. We can adjust the running
time of the algorithm arbitrarily by a factor for the initial and
finial Hamiltonians of the adiabatic evolution. Especially, we can
reduce the running time to a constant time independent of the size
of the database. In Grover's algorithm, we can't do this.

To begin, we briefly depict the quantum adiabatic theorem and
quantum search algorithm by local adiabatic evolution for later
convenience (see \cite{RC02} for details). Suppose a quantum
system with a time dependent Hamiltonian $H(t)$ is in a state
$|\psi(t)\rangle$. The quantum adiabatic theorem says that if the
system is initially in the ground state of $H(0)$ and the
Hamiltonian varies slowly enough (Suppose the time of the process
is T, then T is not very little), the state of the system will
stay close to the instantaneous ground state of the Hamiltonian at
each time t.

Concretely, suppose $H_0$ and $H_m$ are the initial and finial
Hamiltonians of the system, and suppose the system is evolved by
the following time dependent Hamiltonian:
$$ H(t)=(1-s)H_0+sH_m, $$
where $s=s(t)$ is a monotonic function with $s(0)=0 $ and
$s(T)=1$. Let $|E_0,t\rangle$ and $|E_1,t\rangle$ be the ground
state and the first excited state of the Hamiltonian at time t,
then let $E_0(t)$ and $E_1(t)$ be the corresponding eigenvalues.
By the adiabatic theorem we have
$$|\langle E_0;T|\psi(T)\rangle|^{2}\geq1-\varepsilon^2$$
provided that
\begin{equation}
\frac{D_{max}}{g_{min}^2}\leq\varepsilon,\ \ \ \ \varepsilon\ll1,
\end{equation}
where $g_{min}$ is the minimum gap between $E_0(t)$ and $E_1(t)$
$$g_{min}=\min_{0\leq t \leq T}[E_1(t)-E_0(t)],$$
and $D_{max}$ is a measurement of the evolving rate of the
Hamiltonian
$$D_{max}=\max_{0\leq t \leq T}|\langle\frac{dH}{dt}\rangle_{1,0}|.$$

Roland and Cerf in \cite{RC02} showed how to implement quantum
search by local adiabatic evolution. We can do as follows. Let
$$H_0=I-|\psi_0\rangle\langle\psi_0|,$$
$$H_m=I-|m\rangle\langle m|,$$
where
$$|\psi_0\rangle=\frac{1}{\sqrt{N}}\sum\limits_{x=1}^{N}|x\rangle.$$
$|\psi_0\rangle$, the ground state of $H_0$, is the initial state
of the system just as in Grover's algorithm, and $|m\rangle$, the
destination state of the evolution, is the ground state of $H_m$.
If we evolve the Hamiltonian carefully, after a proper time T we
can get the right solution almost determinately by measurement. It
is shown that the lower bound of the running time T is
$$T\simeq\frac{\pi}{2\varepsilon}\sqrt{N},$$
which gives rise to a quadratic speed up in contrast to classical
algorithms just as Grover's algorithm.

In the above algorithm we have no priori probability about the
solution. Suppose we have a priori probability, we use this
information to improve our search algorithm. To state our result
the following theorem is useful.

\begin{Theorem} Suppose we search a unsorted database with N items
for one marked item using the adiabatic evolution. If the initial
state of the algorithm is
$$|\psi_0\rangle=\sum\limits_{x=1}^{N}a_x|x\rangle,   \ \ \   0<a_m\ll1,$$
the running time of the algorithm is
$$T\simeq\frac{1}{a_m}\times\frac{\pi}{2\varepsilon},$$
where $m$ is the only marked item of the database.
\end{Theorem}
{\it Proof.} We let
$$H_0=I-|\psi_0\rangle\langle\psi_0|,$$
$$H_m=I-|m\rangle\langle m|,$$
then
$$H(s)=(1-s)(I-|\psi_0\rangle\langle\psi_0|)+s(I-|m\rangle\langle m|).$$

To estimate the running time of the algorithm, we need to estimate
$D_{max}$ and $g(s)$, where $g(s)$ is the gap between $E_1(t)$ and
$E_0(t)$
$$g(s)=E_1(t)-E_0(t).$$
It is not easy to calculate the eigenvalues of $H(s)$ in the
computational basis, and we use the orthonormal basis
$\{|\alpha_i\rangle, 1\leq i\leq N\}$ to eliminate the difficulty,
where
$$|\alpha_1\rangle=|m\rangle,$$
$$|\alpha_2\rangle=\frac{1}{\sqrt{1-a^2_m}}(|\psi_0\rangle-a_m|m\rangle),$$
and we don't need to care $|\alpha_i\rangle$, $3\leq i\leq N$.

Then we have
$$|\psi_0\rangle=\sqrt{1-a^2_m}|\alpha_2\rangle+a_m|\alpha_1\rangle.$$
Now it is not difficult to check that in the new orthonormal basis
$H(s)$ will be
$$H(s)=
\begin{pmatrix}
a^2_m(s-1)-s+1 & a_m\sqrt{1-a^2_m}(s-1) \\
a_m\sqrt{1-a^2_m}(s-1) & (1-a^2_m)(s-1)+1\\
& & I\\
\end{pmatrix},$$
where the empty spaces of the matrix are all zeroes. It is easy to
get the eigenvalues of $H(s)$, we denote them by
$\lambda_i$,$1\leq i \leq N$, then
$$\lambda_1=\frac{1}{2}-\frac{1}{2}\sqrt{1-4(1-a^2_m)s(1-s)},$$
$$\lambda_2=\frac{1}{2}+\frac{1}{2}\sqrt{1-4(1-a^2_m)s(1-s)},$$
and $\lambda_i=1$ for $3\leq i\leq N$.

So, we have
$$g(s)=\sqrt{1-4(1-a^2_m)s(1-s)}.$$
On the other hand, It turns out that for arbitrary state
$|\alpha\rangle$ we find
$$\langle E_1,t|\alpha\rangle\langle\alpha|E_0,t\rangle\leq \frac{1}{2},$$
which gives
$$|\langle\frac{dH}{ds}\rangle_{1,0}|=|\langle E_1,t|(|\psi_0\rangle\langle\psi_0|-|m\rangle\langle m|)|E_0,t\rangle|\leq 1.$$
According to the adiabatic theorem, the system must evolve under
the following condition:
$$|\frac{ds}{dt}|\leq \varepsilon \frac{g^2(s)}{|\langle\frac{dH}{ds}\rangle_{1,0}|}.$$
To make the evolution as fast as possible, we can let $s(t)$
satisfies the equation
$$\frac{ds}{dt}=\varepsilon g^2(s) = \varepsilon [1-4(1-a^2_m)s(1-s)].$$
By integral, we can get the lower bound of the running time of the
whole evolution
$$T=\frac{1}{\varepsilon}\times\frac{1}{a_m\sqrt{1-a^2_m}}\arctan\frac{\sqrt{1-a^2_m}}{a_m}.$$
Because $a_m\ll1$, then we have
$$T\simeq\frac{1}{a_m}\times\frac{\pi}{2\varepsilon}.$$
That completes the proof of this theorem. \hfill $\Box$

We can see that the running time is affected only by $a_m$, and is
independent of $a_x,x=1,\dots,N,x\neq m$. We can try to think
about this property in the standard quantum computation paradigm.
In Grover's algorithm, it is difficult to calculate the accurate
running time if the initial state is very disordered. Furthermore,
the running time of Grover's algorithm is affected by all the
amplitudes of the initial state.

Now we use the above result to deal with the search problem under
a priori probability. About this problem, we have the following
theorem.

\begin{Theorem} Suppose we search a unsorted database with N items for a
solution using the adiabatic evolution. Before search we have
gotten a priori probability,
$$P(m\in A_i)=p_i\neq 0,\ \ \sum\limits_{i=1}^{k}p_i=1,\ \ 1\leq i\leq k.$$
Where ${A_i}$ is a partition of the set $\{1,\dots,N\}$,

$$\cup_{i=1}^{k}A_i=\{1,\dots,N\}, \ A_a\cap A_b=\phi,\ 1\leq a< b\leq k.$$
Then we can improve the performance of search by the priori
probability, the new running time is
$$T\simeq \sqrt{\frac{n_M}{p_M}}\times \frac{\pi}{2\varepsilon},$$
where $n_i$ is the number of the items in $A_i$, and $M$ is the
index of the subset that contains the marked item $m$. The mean
running time of the modified algorithm is less than that of
original adiabatic search algorithm. For convenience we suppose
$p_i/n_i\ll 1$ for $1\leq i\leq k$.
\end{Theorem}

{\it Proof.} The proof is easy if we notice that before search we
don't know the location of the marked item and have to set the
initial amplitudes averagely in every subset according to the
priori probability of the subset. By theorem 1, it can be show
that
$$T\simeq \sqrt{\frac{n_M}{p_M}}\times \frac{\pi}{2\varepsilon}.$$
The mean running time of the algorithm is
$$T_{mean}=\sum_{i=1}^{k}p_i\sqrt{\frac{n_i}{p_i}}\times \frac{\pi}{2\varepsilon}.$$
Using Cauchy-Schwartz Inequality, it is easy to prove
$$T_{mean}\leq \frac{\pi}{2\varepsilon}\sqrt{N},$$
where the quantity appearing on the right hand side of the
inequality is the running time of adiabatic search algorithm
without priori probability. The equality holds if and only if
every subset has proportional priori probability. \hfill $\Box$

Now we use some instances to show the effect of Theorem 2. Suppose
the partition contains only one subset, the search problem
essentially degenerates to the instance without priori
probability. It can be shown that the running time is equal to the
result in \cite{RC02}. While If the partition contains two subsets
$A_1=\{1,\dots,N/2-1\}$ and $A_2=\{N/2,\dots,N\}$ with priori
probability $p_1=0.8$ and $p_2=0.2$ respectively, we can find that
the running time by Theorem 2 is about 20 percent faster than the
result in \cite{RC02}.

By Theorem 2, we have shown that the priori probability about the
marked item makes the search performance better. From above we
also find that it is convenient to calculate the accurate
performance of adiabatic quantum search algorithm with a
complicated initial state, while this task is not easy in Grover's
algorithm. Besides, adiabatic algorithm has other properties that
conventional quantum algorithms don't have. Theorem 3 is one of
them.

\begin{Theorem}Suppose $H_1(s)$ and $H_2(s)$ are the Hamiltonians of two adiabatic
algorithm, and $T_1$ and $T_2$ are the running time respectively.
If $H_2(s)=cH_1(s)$ for $0\leq s\leq 1$, then
$$T_2=\frac{1}{c}T_1,$$
where $c>0$.
\end{Theorem}
{\it Proof.} We notice that if the Hamiltonian is multiplied by a
factor, the eigenvectors don't change and every eigenvalue changes
by the same factor. Thus the proof is not difficult.   \hfill
$\Box$

In Grover's algorithm, we can't adjust the running time of the
algorithm arbitrarily, but in the adiabatic search algorithm we
can do this by a factor for the Hamiltonian. Furthermore, this
property can be extended to other adiabatic quantum algorithms. So
this property can give convenience to us if in the future we use
adiabatic evolution to perform some tasks.

To our surprise, we can reduce the running time of adiabatic
search algorithm to a constant time independent of the size of the
database if we use a proper factor. For example, in \cite{RC02} if
we set the initial and finial Hamiltonians as follows,
$$H_0=\sqrt{N}I-\sqrt{N}|\psi_0\rangle\langle\psi_0|,$$
$$H_m=\sqrt{N}I-\sqrt{N}|m\rangle\langle m|,$$
the running time will change to
$$T\simeq\frac{\pi}{2\varepsilon}.$$
It is a constant time.

In conclusion, we have shown two properties of the adiabatic
search algorithm that don't exhibit in the corresponding algorithm
of the standard paradigm. By the first one, we have shown that if
we search a database under a priori probability, we can use the
priori probability to improve the performance. Moreover, the
running time of adiabatic evolution can be changed arbitrarily by
a factor for the Hamiltonians. It seems that sometimes it is more
convenient to deal with quantum algorithms by adiabatic evolution
than by quantum circuit. In fact, by adiabatic evolution several
novel quantum algorithms have been proposed
\cite{FGG01}\cite{TDK01}.


\begin{thebibliography}{9}
\bibitem{GROVER97} L. K. Grover, Phys. Rev. Lett 79, 325(1997).

\bibitem{FGGS00} E. Farhi, J. Goldstone, S. Gutmann, and M. Sipser, e-print quant-ph/0001106.

\bibitem{RC02} J. Roland and N. J. Cerf, Phys. Rev. A 65, 042308(2002).

\bibitem{FGG01} E. Farhi et al. e-print quant-ph/0104129.

\bibitem{TDK01} T. D. Kieu, e-print quant-ph/0110136.

\end{thebibliography}
\end{document}